\documentclass[12pt]{iopart}
\usepackage{cite}  
\usepackage[dvips]{graphics}
\usepackage{epsfig}

\begin{document}

\title[]{Effects of Laser Wavelength and Density Scalelength on Absorption of Ultrashort Intense Lasers on Solid-Density Targets}

\author{Susumu Kato\footnote[3]{(s.kato@aist.go.jp)}, Eiichi Takahashi, Tatsuya Aota,\\
Yuji Matsumoto, Isao Okuda,  and Yoshiro Owadano}

\address{National Institute of Advanced Industrial Science and Technology (AIST), Tsukuba, Ibaraki 305-8568, Japan}

\begin{abstract}
Hot electron temperatures and electron energy spectra in the course of interaction between intense laser pulse and overdense plasmas are reexamined from a viewpoint of the
difference in laser wavelength. The hot electron temperature measured by a particle-in-cell simulation is scaled by $I$ rather than $I\lambda^2$ at the interaction with overdense plasmas with fixed ions, where $I$ and $\lambda$ are the laser intensity and wavelength, respectively.
\end{abstract}





\section{Introduction}
The interaction of intense laser pulses with overdense plasmas has attracted much interest for the fast ignitor concept in inertial fusion energy \cite{Tabak1994}. The interaction of ultrashort intense laser pulses with thin solid targets have also been of great interest for the application to high energy ion sources \cite{ionsource}. Ultraintense irradiation experiments using an infrared subpicosecond laser, e.g., Nd:glass ($\lambda =$ 1,053 nm) or Ti:sapphire ($\lambda =$800 nm) lasers, whose powers and focused intensities exceed 100 TW and $10^{20}$ W/cm$^2$, are possible using chirped pulse amplification techniques \cite{Maine:1988}. In these experiments, the classical normalized momentum of electrons $a_0\equiv
P_{\rm osc}/mc=({I\lambda_\mu^2}/{1.37\times10^{18}})^{1/2} \geq 1$, where $m$ is the electron mass, $c$ is the speed of light, $I$ is the laser intensity in W/cm$^2$, and $\lambda_\mu$ is the wavelength in $\mu$m. On the other hand, a KrF laser  ($\lambda =$ 248 nm) has an advantage as the fast ignitor in that the critical density is close to the core, and hot electron energies are suitable since the critical density of the KrF laser is ten times greater than that of an infrared laser \cite{Shaw:1999}. The peak intensities of KrF laser systems were only the order of $10^{18}$ W/cm$^2$, namely $a_0 < 1$ \cite{Teubner:1996}. Therefore, the dependence of the laser plasma interactions on the laser
wavelength was not investigated in $a_0 \geq 1$. Recently, the laser absorption and hot electron generation have been studied by the high intensity KrF laser system of which focused intensity is greater than $10^{19}$ W/cm$^2$ \cite{takahashi2004}. However, the production of hot electrons by the high intensity KrF laser has not been fully understood yet. Namely, it has been not clear that the effects of laser wavelength on hot electrons produced by ultrashort intense laser pulse on solid-density targets.

The absorption, electron energy spectrum, and hot electron temperature have usually been investigated and scaled using the parameters $I\lambda^2$, $n_{\rm{e}}/n_{\rm{c}}$, and $L/\lambda$ \cite{Lefebvre:1997,Wilks:1997}, where $n_{\rm{e}}$, $n_{\rm{c}}$, and $L$ are the electron density, critical density, and density scale length, respectively. 
Critical density absorption of the laser light converts laser energy into hot electrons
having a suprathermal temperature $T_{\rm {h}}$ approximately
proportional to $\sqrt{I\lambda^2}$ for $a_0 > 1$, and $T_{\rm{h}}\sim
[(1+a_0^2)^{1/2}-1]mc^2$ at moderate densities \cite{Wilks:1992}, where
$mc^2=511$ keV, $m$ is an electron rest mass.
The scaling of the hot electron temperature has been supported by experiments
 of Nd:glass and Ti:sapphire lasers \cite{Malka:1996}. 
On the other hand, the results of one-dimensional simulation for
 normal incidence in the density region $4 < n_{\rm{e}}/n_{\rm{c}} <
 100$ and the normalized intensity $4 < a_0^2 < 30$ have shown that 
$T_{\rm{h}} \sim \eta \left(n_{\rm{e}}/n_{\rm{c}}\right)^{\alpha} 
[(1+a_{\rm s}^2)^{1/2}-1] mc^2$, where $a_{\rm
 s}= \beta a_0\left(n_{\rm{c}}/n_{\rm{e}}\right)^{1/2}$ is the electromagnetic
 fields at the surface of the overdense plasma, $\eta = 0.5\sim1.1$ and
 $\alpha = 1/2$, which depend weakly on $I\lambda^2$ and
 $n_{\rm{e}}/n_{\rm{c}}$ \cite{Wilks:1997}. $\beta$ is weakly depend on the angle of incidence, absorption rate, and $n_{\rm{e0}}/n_{\rm{c}}$\cite{Lichters1996}.
 The hot electron temperature is scaled by the amplitude of
 electromagnetic fields at the plasma surface rather than that in
vacuum; namely, the hot electron temperature is slightly
dependent on the wavelength.

In addition, at the interaction of intense laser pulses with solid density plasma which has a sharp density gradient the hot electron temperature is scaled by $I\lambda^{-1}$ rather than $I\lambda^2$ \cite{kato2002}. 
In the present paper, we study the absorption of ultrashort intense laser pulses on overdense plasmas for different laser wavelengths ($\lambda$ = 0.25, 0.5, and 1 $\mu$m) using a particle-in-cell (PIC) simulation.

\section{PIC simulation}
In order to investigate hot electron generation for oblique incidence, we use the
relativistic 1 and $2/2$ dimensional PIC simulation with the
boost frame moving with $c \sin \theta$ parallel to the target surface,
where $c$ and $\theta$ are the speed of light and an angle of
incidence\cite{Gibbon:1992}.  
In the simulation, the target is the fully ionized plastic and the electron density $n_{\rm{e0}} \sim 3.5 \times10^{23}{\rm cm^{-3}}$. 
The density correspond to
$n_{\rm{e0}}/n_{\rm{c}}=20$, $78$, and $310$ for $\lambda =$ 0.25, 0.5, and $\lambda = 1 \mu$m, respectively. 
The density profile has a sharp density gradient, $n_{\rm{e}}(x)=
n_{\rm{e0}}$ for $x \ge 0$ and $n_{\rm{e}}(x)= 0$ for $x<0$.  In order to clarify the boundary  effect, ions are fixed, namely, the boundary does not move all the time.
The laser pulse starts at $x < 0$ and propagates towards $x > 0$. The laser
intensity rises in 5 fs and remains constant after that. The irradiated
intensity $I=5\times10^{19}$ W/cm$^2$ and the angle of incidence
$\theta=30^\circ$ and $45^\circ$ (p-polarized), respectively. $a_0^2 =$ 2.3, 9.2, and 36 for
$\lambda =$  0.25, 0.5, and 1.0 $\mu$m, respectively. However, $a_s^2=0.12\beta$ for all wavelength. 
Normalized electron energy distributions after 50 fs are shown in
Fig.1(a) and 1(b) for $\theta=30^\circ$ and $45^\circ$, respectively.
 The hot electron temperatures are 140 and 340 keV for $\theta=30^\circ$ and $45^\circ$, respectively.  The hot electron temperatures does not depend on the laser wavelength. The result is well agreement with that of a simple sharp boundary theory. On the other hand, the absorption depends on the laser wavelength, $A(\theta=30^\circ)=$ 0.9-1.8\%, 2.2-3.0\%, and 3.6-4.3\% and $A(\theta=45^\circ)=$ 2.6-4.1\%, 5.3-6.7\%, and 7.8-9.0\% for $\lambda =$ 1.0, 0.5, and 0.25 $\mu$m, respectively.

 \section{Concluding Remarks}
The effects of laser wavelength on hot electrons produced by ultrashort intense laser pulse on solid-density targets are studied by the use of a PIC simulation.
As a result, the dependence to the wavelength of hot electron temperature strongly depend on the boundary condition, even in the one dimensional case, namely, all are not determined only by $I\lambda^2$.
The density profiles of both preformed plasma \cite{Yu:2000} and multi-dimensional effects such as surface deformation \cite{Wilks:1992} are very important in the actual experiments.

 \ack {A part of this study was financially supported by the Budget for Nuclear Research of the Ministry of Education, Culture, Sports, Science and Technology, based on the screening and counseling by the Atomic Energy Commission.}

\newpage
\begin{figure}[htbp]
\begin{center}
 \includegraphics[width=80mm]{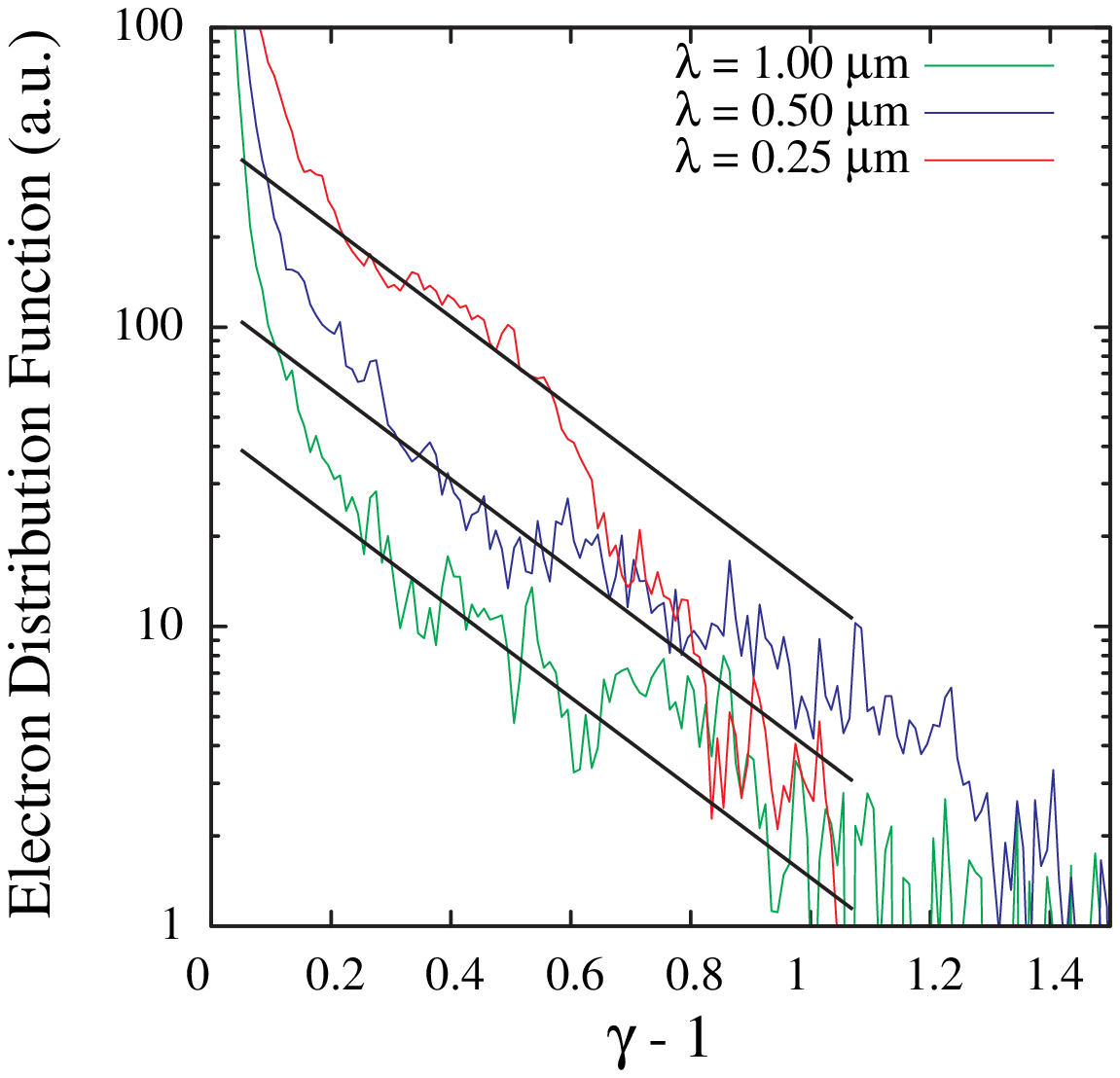}
\end{center}
\vspace{1cm}
\begin{center}
 \includegraphics[width=80mm]{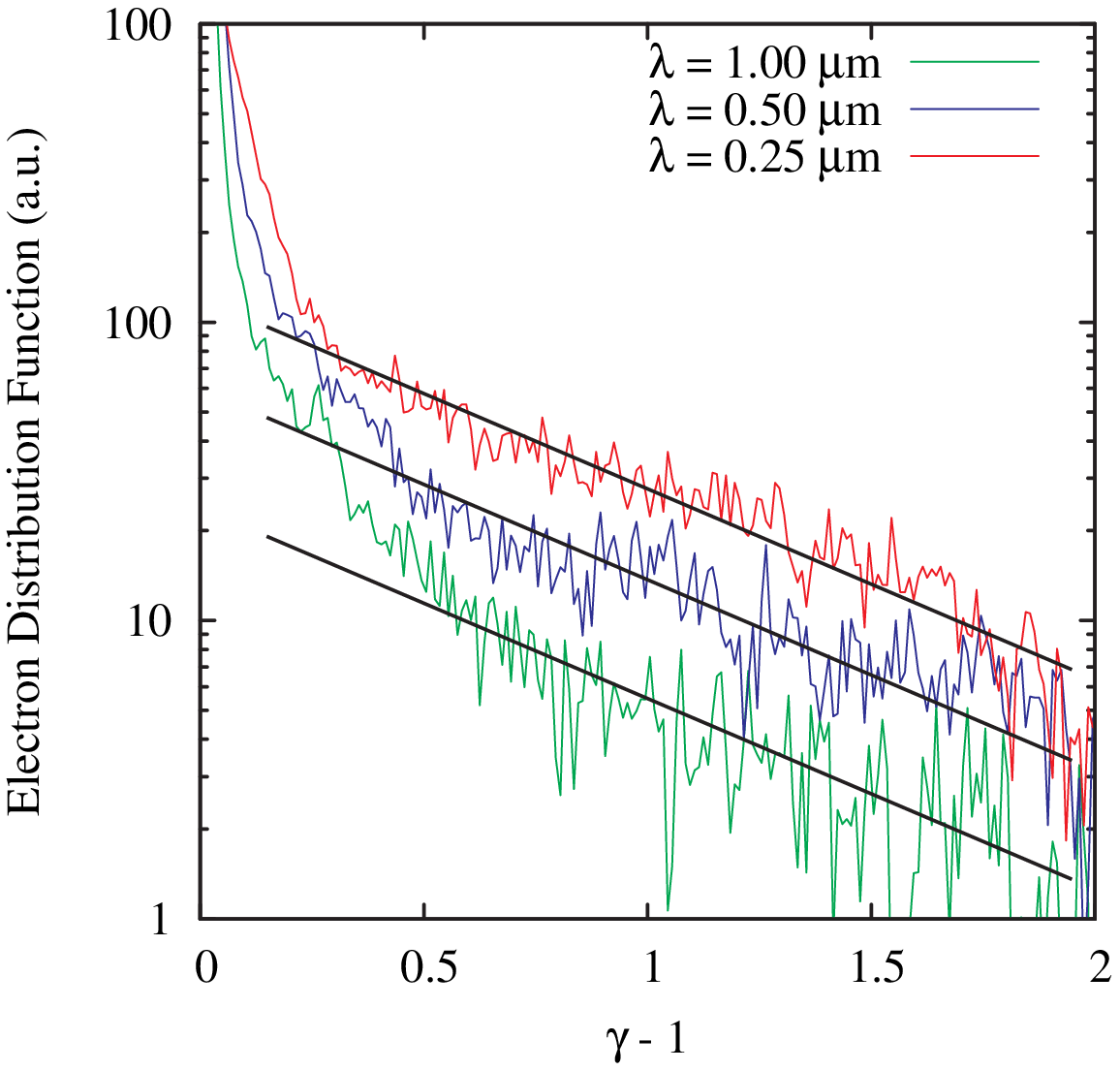}
\end{center}
\end{figure}
\Figure{\label{label1} Electron energy distribution at t $=$ 50 fs and $I=5\times10^{19}$
 W/cm$^2$ for (a) $\theta=30^\circ$ and (b)$\theta=45^\circ$, respectively. The red, blue, and green lines are for $\lambda$ = 0.25, 0.5, and 1 $\mu$m, respectively.}

\end{document}